# Can Online MBA Programs Allow Professional Working Mothers to Balance Work, Family, and Career Progression? A Case Study in China

Mboni Kibelloh, Yukun Bao[*]

Centre for Modern Information Management,

School of Management, Huazhong University of Science and Technology, Wuhan, P.R.China, 430074

**Abstract**

Career progression is a general concern of professional working mothers in China. The purpose of this paper is to report a qualitative study of Chinese professional working mothers that explored the perceptions of online Master's of Business Administration (MBA) programmes as a tool for career progression for working mothers balancing work and family in China. The objective was to examine existing work-family and career progression conflicts, the perceived usefulness of online MBA in balancing work-family and career aspirations, and the perceived ease of use of e-learning. Using Davis's (1989) technology acceptance model (TAM), the research drew on in-depth interviews with 10 female part-time MBA students from a university in Wuhan. The data were analysed through coding and transcribing. The findings showed that conflicts arose where demanding work schedules competed with family obligations, studies, and caring for children and the elderly. Online MBA programmes were viewed as a viable tool for balancing work and family and studying, given its flexible time management capabilities. However, consideration must be given to address students' motivation issues, lack of networking, lack of face-to-face interaction, and quality. The research findings emphasise the pragmatic need to re-align higher education policy and practice to position higher education e-learning as a trustable education delivery channel in China. By shedding light on the prevailing work-family conflict experienced by women seeking career advancement, this study suggests developing better gender-supporting policies and innovative e-learning practices to champion online MBA programme for this target niche.

[*] Corresponding author: Tel: +86-27-87558579; fax: +86-27-87556437.
Email: yukunbao@hust.edu.cn or y.bao@ieee.org

# 1. Introduction

Career progression is a general concern of professional working mothers in China. Extensive research abounds relating work-family conflict to the under-representation of women in top management (Blair-Loy, 2003). Still to be addressed, however, is how women can realise individual agency in developing and achieving career progression (Broadbridge & Simpson, 2011). One potential route to more senior management positions for women is the pursuit of management education through a Master's of Business Administration degree (MBA) (Finney, 1996; Simpson, 2000). An MBA is considered an important pre-requisite for both men and women who aspire to senior positions (Finney, 1996); an effective tool against gender discrimination (Leeming & Baruch, 1998); and a passport to fast-track career and senior managerial roles (Baruch & Peiperl, 1999). Although the benefits of an MBA for career progression are well-documented in the literature from a non-Chinese perspective, little empirical evidence exists in the context of China's economy, especially for the career development of women (Chen et al., 2012).

Conversely, these MBA students already lead complex lives as employees, employers, spouses, mothers, community volunteers, and caregivers for elderly and ill relatives (Crosby, 1991). An alternative to the emotional rigours of trying to balance work, family, and studies is to opt for online education. Many researchers have suggested that online education (also referred to as e-learning) is a viable option for busy and working mothers (Home, 1998), offering learning opportunities 'anywhere' and 'anytime'. However, these changing patterns in education have been found to induce anxiety and uncertainty in users (e.g., Ong & Lai, 2006), and many conceptual boundaries (e.g., technological, pedagogical, social, economic) have yet to be fully understood or explored (Rossiter & Crock, 2006). In particular, the conceptual boundaries of the work-family conflict, career aspirations, and the overall impact on online MBA acceptance for women in China remain relatively unexplored.

Additionally, online education is still a new concept for many people in China, and attitudes towards its adoption have not been fully studied (Duan et al, 2010). A better understanding of online education adoption intentions in China, particularly those of professional working mothers, would enable providers to offer courses that are more likely to be utilised by future online learners. This research aims to examine online MBA programme adoption intentions in China from a technology acceptance perspective.

To achieve the research objectives, Davis's (1989) technology acceptance model (TAM) was employed as one of the most popular theoretical frameworks for predicting system acceptance of technology. In this study, the TAM model is used to investigate the perceived usefulness and ease of use of online MBAs in a qualitative case study by interviewing 10 female professional part-time MBA students.

This case study shows that female MBA students' expectations may present challenges for educational practitioners, e-learning developers, and policy makers intending to exploit the flexibility of online MBAs in China. This research makes a number of contributions to both theory and practice by applying the TAM model to a qualitative case study in the context of China. The authors propose that a number of key issues need to be considered carefully when promoting and supporting e-learning and the use of online MBAs for career progression. Furthermore, a range of interventions is needed to re-align educational policy and practice with prevailing labour market requirements.

This paper provides a background for the work-family conflict and career progression followed by a literature review of the use of MBAs for career progression and online MBAs. A theoretical explanation of the TAM theory and description of the methodology provides synthesised findings leading up to a discussion of the results, their implications, and suggestions for future practice and research, which concludes the study.

## 2. Work-family conflict and career progression

In China, women make up to 38% of the fulltime work force and are overrepresented in manufacturing, services, and public sector industries such as health, education, and social welfare (Cooke, 2012). However, their presence in managerial levels reflects a gender gap in terms of career progression (Cao, 2001) and salary increments (Shu & Bian, 2003). Prior research relates the failure of women to progress to senior management with work-family conflict (Broadbridge & Simpson, 2011; Blair-Loy, 2003). Greenhaus & Beutell (1985, p. 77) define work-family conflict as: "a form of inter-role conflict in which the role pressures from the work and family domains are mutually incompatible in some respect".

The ideal worker norm has long been associated with men, who are expected to devote more time to work. Women who try to fit into this norm and serve as primary caregiver may find it difficult to balance the demands of both an ideal worker and an ideal caregiver (Williams, 2005). Women's work and childbearing lifecycle patterns are diametrically opposed to the senior management career lifecycle, where the intensive workload and commitment necessary to succeed coincide with peak child-rearing years (Drew & Murtagh, 2005). This difficulty then becomes the primary source of women's disadvantage in the corporate world and explains their "concentration in low paid, part-time employment and their absence at the most senior levels of management" (Doherty, 2004) (p. 433).

The one-child policy in China has meant that the only-child generation is more precious than the previous generations of children, and parents, particularly those in the middle class, compete against each other in bringing up their prodigy (Xiao & Cooke, 2012). Children are expected to start their serious education well before the official schooling age of six. Therefore, despite being able to afford paid childcare, middle-class mothers may be under pressure to channel their energy into developing their only child instead of their own career (Xiao & Cooke, 2012).

Although it is customary for Chinese women to receive childcare support from their live-in parents, caring for elderly parents, especially if ill or particularly old, is an additional family demand. Managing elder care has shown to be more complex than managing childcare because it involves the coordination of many social services (Friedman & Galinsky, 1992). Subsequently, even the strong support typical of Chinese families and social groups does not appear sufficient to alleviate women's childcare and domestic burdens, nor does it allow them to overcome cultural barriers to their career aspirations, resulting to a failure to achieve top positions in management (Ng et al, 2002).

# 3. Literature review and theoretical framework

## 3.1 MBA for career progression

Obtaining a good job in China still depends upon obtaining advanced education and using one's guanxi *(network)* to persuade a personnel manager to hire oneself (Granrose, 2005). Guanxi is a complex web of social connections and mutual obligations used to exchange favours and conduct business in Chinese society (Park and Luo, 2001). In Hong Kong, higher levels of education have been associated with higher income and more prestigious careers (Cheng & Yuen, 2012). Similarly, several studies have found significant improvements in the career progression of managers after completing an MBA course (Association of MBAs, 1992). Although some (e.g., Mintzberg, 2004) have argued against MBA in developing interpersonal skills required for effective management and leadership, Simpson (2000) found that an MBA increased confidence and credibility, providing information collection and analysis, quantitative analysis, technology management, entrepreneurial, and action skills (Boyatzis & Case, 1989). Nonetheless, an MBA serves as an effective means of acquiring managerial competencies and enhancing career prospects (Finney, 1996). An MBA can therefore be considered an important component of a Chinese professional women's career progression. However, the inter-role conflicts experienced by women working a double or even triple shift, including career, children, ageing parents, and study, necessitates flexible learning delivery channels, such as online education and an online MBA.

## 3.2 The technology acceptance model (TAM)

This study employs Davis's (1989) technology acceptance model (TAM) as the theoretical grounding for exploring factors influencing the perception of online MBA programmes. TAM, adapted from the theory of reasoned action (TRA) (Fishbein & Ajzen, 1975), has been used as the theoretical basis for many empirical studies of user technology acceptance (Davis, 1989; Ong & Lai, 2006; Teo et al, 2012; Venkatesh &

Morris, 2000). This model is perhaps the most promising direction for attempts to overcome the problem of underutilised systems. Unfortunately, there is little evidence of TAM being applied to professional working mothers in the context of China.

The TAM model is comprised of two prominent variables, perceived usefulness and perceived ease of use. Perceived usefulness has been defined as an indicator of the extent to which a person believes that using a particular technology will enhance his or her performance and therefore represents an individual's extrinsic motivation to use a technology (Davis, 1989). A significant body of prior research has shown that perceived usefulness has a positive effect on behavioural intention to use (Davis et al, 1989; Venkatesh & Morris, 2000). Conversely, perceived ease of use refers to the degree to which a person believes that the use of a particular technology will be free of effort and is therefore an indicator of an individual's intrinsic motivation to use a technology (Davis, 1989). Venkatesh & Morris (2000) found that a low evaluation of perceived ease of use caused an increase in the salience of such a perception in determining perceived usefulness and user acceptance decisions. In TAM, beliefs that a technology is useful and easy to use influence the users' attitudes toward the technology and thereby their decision to adopt the technology. The need to balance work-family commitments and career aspirations would likely position online MBA programmes as a useful alternative to the traditional classroom education delivery method and is expected to influence the decision to pursue an online MBA.

## 3.3 E-learning and online MBA programmes

The greatest attraction of online MBA programmes is the convenience and flexibility of the delivery channel. Online education enables adults with full-time jobs to attend classes without having to leave their current jobs (Lorenzo, 2004). This allows students to maintain employment and other family responsibilities while being able to conveniently continue with their education with a flexible schedule and low travel costs and enables students to interact with teachers and students from around the world (Hung et al, 2010).

Studies have found student satisfaction and perceived usefulness to be key factors in explaining learners' behavioural intention to use e-learning (Liaw, 2008). Lee, Yoon, & Lee (2009) presented the significant influence of instructor characteristics and teaching materials on the perceived usefulness of e-learning, while perceived usefulness and enjoyment were predictors of the intention to use e-learning. On the other hand, Visser, Plomp, & Kuiper (1999) showed that the student's characteristics and motivation to learn predicted participation in e-learning. Furthermore, research has shown that perceived networking, instructor interaction with students, quality, and active discussion (Swan et al, 2000, Jiang & Ting, 1998) have a significant impact on perceptions of online education.

## 3.4 Research questions

This study aims to answer the following research questions (RQs):

*RQ1-* What are Chinese professional women's experiences with work-family conflict and career aspirations?

*RQ2-* What is the perceived usefulness of an online MBA programme in terms of balancing work-family demands and career aspirations?

*RQ3* –What is the perception of e-learning in terms of ease of use?

# 4. Methodology

## 4.1 Research method

Although TAM has been the subject of investigation of a large number of studies, many such studies are limited in several respects, such as the strictly positivist quantitative perspective of research focusing on the adoption of technologies as such (Davis, 1989). This study helps to address this limitation in the literature by providing an in-depth qualitative study. This decision is in accordance with recommendations of proponents of the case study approach, such as Yin (1994). Given the complexity of the work-family and career progression conflict, the authors found it necessary to record the informants' experiences and thoughts regarding online MBA programmes instead of using structured questionnaires, which would have risked omitting critical information.

## 4.2 Data collection

In-depth face-to-face interviews were conducted in the spring of 2012 with 10 part-time MBA students from a university in Wuhan, China. Demographic data including age, marital status, and education and career background; patterns of career progression; and availability of home and child support were first collected using semi-structured interviews. Next, the interviewer asked three open-ended questions concerning the informant's perception of the experiences of work-family conflict and career progression; the usefulness of online MBA programmes in helping to balance work-family and career progression; and the ease of use of e-learning technology.

The snowball sampling method was used to recruit female participants. This method relies on referrals from the initial subjects to generate additional subjects. Interview invitations with criteria (working mothers) were sent to part-time MBA students via the class monitor. The initial participants were then encouraged to bring along fellow classmates with similar backgrounds. This method was suitable and effective for recruiting the appropriate target group because student working mothers knew other student working mothers.

The average age of the informants was 28 years old. Their backgrounds included automobile (2 participants), electrical (3 participants), software applications (2 participants), human resources (1 participant), and retail (2 participants). Interview questions were given to participants prior to the interview, and participants were guaranteed the confidentiality of their information to ensure they spoke freely. The interviews were audiotaped, and notes were taken to ensure accurate recording of the responses and the interviewer's overall impressions. To protect the participants' identities, pseudonyms are used in this report. All interviews were conducted in English and generally lasted approximately one hour each. The data were then sorted into a database manually by one of the authors and checked by the other author.

## 4.3 Data analysis

Theoretical thematic data analysis was adopted to analyse the case data, following Braun & Clarke's (2006) qualitative data analysis model. First, the researchers read through the transcripts and jotted down comments, notes, thoughts, and observations in the margins. Next, the researchers went over the marginal notes to summarise key issues and to section and categorise the data. Code labels were assigned to each section using the interviewee's words or the researchers' own words.

The preliminary codes were examined for overlaps and redundancy. Eliminating redundant codes and collapsing similar codes enabled the codes constructed in the early stage to be narrowed down into broader themes. The new list of code words was then used to examine the texts to check whether these codes recorded common themes and recurring patterns. The different data sets were continuously read and analysed to refine the categories and to make sure that no text sections were overlooked.

During the analysis phase, the researcher continuously linked the recurring themes to TAM as the theoretical lens of this study. TAM was used both to organise the categories and as an analytical tool to form an in-depth view of the conceptual meanings of the category under the framework. The themes fell into two main categories: those related to work-family conflict and three sub-categories related to attitudes towards online MBA acceptance. The different components of the TAM model (i.e., perceived usefulness and ease of use) were used as "containers" for arranging data themes (Barab, Evans, & Baek, 2003).

As the interviewers gathered more data and the coding continued, it was found that no new themes were being identified and no new issues arose. Therefore, as suggested by Strauss and Corbin (1990), this study had reached its saturation point with 10 interviews.

To ensure reliability and internal validity, both researchers were involved in the analysis. Where necessary, the informants were contacted for clarification or additional information. For this reason, 2 follow-up telephone interviews were conducted. In addition, peer debriefing was employed. A professional peer who was not directly involved in the data collection but was familiar with the socio-cultural elements of qualitative case study analysis was invited to comment on the findings as they emerged and to check for misinterpretations and researcher bias.

# 5. Findings

## 5.1 Work-family conflict and career aspiration experiences

Many women indicated that work demands and a desire to reach top management levels often required them to focus more on the job than family, at least in the very beginning of employment. This perspective is illustrated below:

> When you want to climb a career ladder, at the first stage, when you get promoted, for at least half a year, you cannot and don't have time for family. You must focus on the job, on the promotion, on the new position. – *Jessica, 36, Program manager*

> In China, the pace of professional life is very fast. When you get promoted from one position to another, time, knowledge, and a lot of work is required to gain experience. There is almost no time to balance family with work; the latter is your only focus. – *Lauren, 27, Market analyst*

Competing with work demands is the importance of investing in childcare and raising children. One informant stated:

> We Chinese put many energy into taking care of our children; we sacrifice everything for our children. – *Veronica, 25, Program manager*

Consequently, managing work, family, and studies was observed to impact women's physical well-being, leaving them tired at the end of the day. One woman stated:

I think it's too tiring and difficult. During work, you have to work and put more effort into working. When you are expecting a baby, it's easy to feel more tired and want to get some sleep. So I think it's too difficult! – *Daisy, 30, Sales manager*

Furthermore, China's collective, familial culture expects the elderly to be cared for and to assist with childcare. The informants expressed the following:

In most of the Eastern or Chinese world, women focus on family, children, and their parents. I think I am no exception. My parents have bad physical health. They are old, weak, and sometimes ill. I put most of my energy, thoughts, and focus on them. So, I need to spend time with them.
– *Veronica, 25, Program manager*

When I observe how my colleagues, co-workers, and classmates cope with their children and work, I think it's very likely that they ask for help from their parents. If they don't, they will have to hire someone to help take care of their children, and that's expensive. And there would still be the worry that she [helper] might not treat my kids like her own. Parents are a great help because they will treat the children like their own, just like their children when they were young.
– *Veronica, 25, Program manager*

## 5.2 Perceived improvement in job prospects with an MBA

Many of the informants believed that an MBA was a ticket to top management positions and higher salaries. The following was stated:

> Some people like me have 8 or 10 years of experience. Eventually, I figured that if I were to enrol in this MBA programme, it would fulfil a large requirement for my promotion…. I have a career development plan and I know I must study more to prepare for this position. Therefore, I believe that after this MBA course, we will have good opportunities…. for career development, we should be ready with experience, and with this programme, we will be ready for promotion.
> – *Jessica, 36, Program manager*

> Salary-wise, you will get at least a 50% raise, which corresponds to 1.5 times your usual pay. This is the first raise, which is very dramatic. I believe that if I keep improving myself, opportunities will present themselves, followed by salary increases. I think that if I instigate the changes myself, good things will come to me. – *Veronica, 25, Program manager*

# 5.3 Women's perceptions of the usefulness of online MBA programmes

Overall the women had positive perceptions of online MBA programmes, centred on the beliefs that online MBA programmes were useful in integrating the role of student with those for work and family. Several themes emerged from the analysis of interview data, which are described below.

### 5.3.1 Flexibility

Many of the women reported flexibility as the most useful aspect of online MBA programmes. Given the need to juggle a full-time job, family, and school, being able to learn at their convenience in terms of time and learning space favourably influenced perception towards online MBA programmes. Examples of relevant comments were as follows.

> I think it's good for us because with e-learning, we can study at home. However, when the children at home ask you to play with them or check their homework, you should comply with their requests and needs, that's the way it is in China. So, when that happens, I cannot study. However, in so many cases, we can use this [e-learning] in an efficient way. – *Jessica, 36, Program manager*

> It has huge benefits, which are more about flexibility and saving time (than anything else), and if you have some urgent issues, you can just press "pause" and continue later. That's great! – *Rebecca, 33, Project manager*

## 5.3.2 Abundant information

Given that e-learning systems contain large repositories of information, the informants believed that they could access more information online than in the limited time spent in classrooms lectures. One of the women stated:

> Being online, you can access more information than in the classroom because on the internet, you can see information on many topics on the same page, whereas in traditional classroom learning, we can only see what the teacher or classmates around you contribute, so that's the difference. – *Veronica, 25, Program manager*

## 5.4 Perceived limitations of online MBA programmes

Although the women were generally positive about the usefulness of online MBA programmes, they also noted that there were some limitations to online learning. The informants cited perceived challenges with self-study motivation, lack of face-to-face networking, and interactions with instructor and fellow students. These themes are described below:

### 5.4.1 Motivation

Some of the informants felt that e-learning required more motivation and self-discipline to allocate time for learning and to sit through an online class compared to classroom learning. One of the women stated:

> Sometimes, when I am online for the programme, I have hard time with self-discipline… I sleep or eat instead, but when you are in a classroom with face-to-face interaction, being focused and concentrating aren't tough! – *Diana, 24, Human resource personnel*

> If you can't see the people, just hear their voices, that's not interesting, it's boring (laughs). If there are pictures online, I can focus my attention on them. In this way, I feel more interested and can study easily. – *Daisy, 30, Sales manager*

### 5.4.2 Lack of face-to-face interaction

Most informants reported that a lack of face-to-face interaction was a major drawback. Some believed that online learning failed to allow them to interact with their instructors and classmates, as in traditional classroom settings. They argued that Chinese students are accustomed to discussing, learning, and working in teams and feel the need to interact closely with the teacher.

Another concern brought up by the informants was the lack of emotional connection between the instructor and student as well as among students. The women stated:

> In classroom learning, we can have face-to-face conversations with our teacher, we can discuss with our classmates, and we can conduct firm case studies, but with e-learning, the efficiency can be based on internet quality. Sometimes lines drop or the teacher cannot be seen face-to-face, or when you see him on the screen of your computer, he is in some session and cannot be in direct contact with you. Thus, mutual feelings are missing, and your understanding and input can be minimised. This is different from the traditional approach. Moreover, sometimes we need to have someone to talk with, and e-learning cannot always offer that opportunity. Under such circumstances, my focus will be worse than in the classroom. – *Veronica, 25, Program manager*

> Actually, I don't like online learning because if we talk face-to-face, I can see you. That way, I can understand how you feel, and you might be able to notice whether I understand what you are saying, and that's good. However, with online learning, you just follow the teacher's lessons, the teachers can't see you and you can't see the teachers, so there are no interactions. So, I like face-to-face instruction much more than virtual instruction. – *Daisy, 30, Sales manager*

### 5.4.3 Perceived networking

Another concern noted was the lack of networking in e-learning. The informants argued that one of the main attractions of an MBA programme was the ability to use this platform to network and share ideas with other professionals. They stated:

Before I wanted to be an MBA student, I wanted to make friends who are also earning their MBA, as they have valuable opinions and ideas and we might be able to work together to do something. I wanted to communicate with others. It's very important! However, in an online classroom, you don't have that chance. – *Diana, 24, Human resource personnel*

[With an MBA], you can have a group of very highly qualified people as your friends. I consider that a great resource in China. – *Veronica, 25, Program manager*

Most people who go back to school after working also want to meet people whom they can talk to on a regular basis and share interests with. In contrast, students who are college graduates want higher education and may not have this desire. However, when you have some working experience, you want others to share their experience with you, which is very important! – *Terry, 31, Electrical engineer*

## 5.5 Women's perceptions of the ease of use of e-learning

This study found very little hesitation regarding the ease of use of e-learning. The informants generally perceived online MBA programmes to be easy to use. However, the issue of easy navigation and access to lessons affected this belief. One informant stated:

> As long as we can easily find the lessons, e-learning is easy to rely on and can be put to good use. – *Jessica, 36, Program manager*

## 5.6 Opinions of the quality of online MBA programmes

Another issue found to impact the perception of online MBA programmes was concerns of the assessment and credibility of online education. This is illustrated below:

### 5.6.1 Assessment

Some of the informants perceived online learning as incapable of verifying that the registered e-learner was in fact the actual candidate being examined. For this reason, the informants questioned the reliability and trustworthiness of online education. They stated:

> I think many people don't have enough time, so they get another person to learn for them, which is a disadvantage. For example, if I don't have time and I need to pass my exam, I can call somebody to learn for me (laughs). For instance, we have an e-learning programme in our company, and my boss gets me to do the e-learning activities for him (laughs). – *Anna, 25, Electronics engineer*

Some of the informants were concerned about how teachers could assess online students' understanding or how online students could self-assess their learning progress. One woman stated:

> In the context of e-learning, I think it's a little hard to check the results. In the classroom, the teacher will ask questions and students will answer. Then, the teacher will know how the students are doing. For me, it's hard to assess the student's level after the completion of training. – *Terry, 31, Electrical engineer*

### 5.6.2 Credibility of online MBA programmes

Another theme that surfaced was the issue of credibility and the reputation of online MBAs, especially for employers. The following was stated:

> You know, in China, fraud is common. Even the course, certificate, or degree delivered to you could be falsified. Online learning is invisible. You cannot touch it, so some people will think it's not credible and that you may not be qualified for a job or worthy of a position with it. – *Rebecca, 33, Project manager*

> I think people trust famous brands; they are the ones with some internationally qualified certification. In this university, our school has its MBA education certified from the system and I think even e-learning platform resources can get the same tool. If they got…, people will judge me as somebody who graduated from a famous school, from a very high platform. If people think or feel that you graduated from a good education agency, they judge your qualifications differently. – *Veronica, 25, Program manager*

# 6. Discussion

The findings of this study present several important issues to be discussed. The following section discusses these results as per the research objectives:

# 6.1 RQ1 – What are Chinese professional women's experiences with work-family conflict and career aspirations?

The findings show that balancing the time allotted for different roles was the predominant cause of fatigue and stress. Work and caring for children and the elderly left little time for study or relaxation. As a result, working mothers risk minimising their social responsibilities when stretching their roles, which in turn affects their prospects of progressing their careers. These findings support Ng et al's (2002) finding that marital and familial roles impact women's progression up the organisational ladder. Nonetheless, undertaking an MBA programme may help these women develop skills to progress their careers.

## 6.2 RQ2 – What is the perceived usefulness of an online MBA programme in terms of balancing work-family demands and career aspirations?

Online MBA programmes are vital in allowing women to balance work-family demands with career aspirations providing the necessary flexibility in terms of time and location to enable them to combine demanding work schedules with family life. Indeed, the informants believe that e-learning is a viable tool due to its flexibility, information availability, and time effectiveness. These findings support Home's (1998) research on online education and work-family balance. However, one finding contradicts Romero (2011), who argues that distance learners have poorer time management due to the lack of time structuring experienced by face-to-face students and that e-learning does not help reconcile the conflict between work, family, and studies.

Nevertheless, concerns were raised by the majority of informants in this study over the lack of face-to-face interactions and networking in e-learning, similar to other studies (McGorry, 2002). The cause for dissatisfaction was due to the fact that, in the special case of China, to develop future career opportunities, both men and women have to develop a network of guanxi contacts and obligations, as explained by Granrose (2007). The guanxi (*network*) differs from other networks in that these social ties have a long-term perspective, are slower to develop and dissolve, and involve a deeper sense of obligation and reciprocal loyalty than is usually present in non-Chinese individuals' concepts of network ties. In addition, China's collective society encourages teamwork and togetherness. There appears to be a misconception that e-learning cannot facilitate student and teacher collaboration or teamwork (Intel, 1997). Furthermore, some informants cited concerns regarding self-motivation. McCall (2002) found that self-motivated learners are more likely to succeed in online learning settings.

## 6.3 RQ3 – What is the perception of e-learning in terms of ease of use?

The informants' perceived ease of use contradicted the initial expectations of this study that ease of use would impact attitude towards and acceptance of online MBA programmes. This study contradicts related studies (e.g., Ong & Lai, 2006). The informants expressed general comfort and confidence with computer usage and the ability to conduct e-learning. This could be because the informants were already experienced computer users due to the demands their daily work, and some were familiar with e-learning from job training. Additionally, many of the informants were from engineering and technology backgrounds with computer knowledge from either work or previous higher education. These findings support prior studies that have shown computing experience to be a strong predictor of attitudes toward computers, computer usage (Whitley, 1997), and subsequent adoption.

## 6.4 Perceived quality of online MBA

Investing in online MBA education is believed to bring valuable returns for future careers. Some informants raised concerns over the quality of online MBA programmes related to the authenticity of the platform for testing students' progress. In other words, how do we ensure that the registered student is the same student being tested? This is particularly of interest in the Chinese case, perhaps due to its long history of examinations, where many pre-defined standards have traditionally been set by educational agencies using strict measures to examine individuals' progress.

## 7. Research and practice implications

This study has contributed to our understanding of the implications of online MBA programmes for Chinese professional working mothers in a number of different ways. It has supported the findings of other researchers (e.g., Home, 1998) that online education is a valuable vehicle for career development when balancing work and family. The findings of this study provide an interesting direction for future research regarding the development and delivery of online MBA programmes and remedies for resolving the work-family and study conflict.

## 7.1 Implications for China's business schools

Leading business schools around the world are increasingly providing women-focused centres. Chinese business schools would benefit from catching up with this trend to lure more women into MBA programmes. Chinese and international business schools should consider marketing to this audience the convenience of online MBA programmes in balancing work-family demands with career aspirations, for which more research is warranted.

This study's findings emphasise the importance of building reputable online learning brands to counteract the fears surrounding quality issues in China. Chinese business schools should consider seeking e-learning accreditation and certifications to further enhance the perceived trust of e-learning as benchmarked by famous and reliable online higher education brands, such as Harvard and MIT.

## 7.2 Implications for online MBA agencies and e-learning practitioners

Due to the lack of perceived networking in online education, higher education institutions in China will need to effectively communicate the versatile and interactive possibilities of online MBAs via marketing campaigns, including web chats, forums, webinars (web seminars), and video conferences. More research is needed to address the ways in which professionals can effectively network and maintain contacts online. Business schools may want to appeal to the perceived usefulness and more supportive environment of collaborative uses of the medium as a way to attract more women to their programmes (Ong & Lai, 2006).

Additional research is also needed to investigate ways in which e-learning lessons can simulate the real-world working environment and market labour requirements with a minimal gap. Online education researchers can consider incorporating lessons on business soft skills, which increase confidence and develop networking skills that can add value to women's managerial careers (Chen et al, 2012). However, e-learning agencies need to address the issue of the assessment of e-learners against learning outcomes given the strong history of examinations and fear of fraud in China. It would be interesting to explore whether these factors influence the perceptions of employers when recruiting or promoting online MBA graduates. Future research could also examine the estimated returns of online MBAs verses classroom-taught MBAs in terms of career and salary increment for both men and women. Similar studies could also compare different national contexts.

Further consideration needs to be paid to mitigating the lack of motivation to engage in online learning, which is important given the already constrained time of a working mother. More research is needed to explore various learning styles, considering that Chinese students tend to prefer learning styles as a collective society. Furthermore, e-learning designers need to be creative regarding interactive multimedia functions such as animations if the programmes are to remain viable.

The participants also highlighted the importance of easy to use e-learning systems. E-learning developers will need to consider user-friendly e-learning environments, such as simple navigation, a robust e-learning infrastructure, and an attractive graphical user interface (GUI). In addition, the availability of support services, such as help desk facilities and troubleshooting, go-to manuals, and 'frequently asked questions' (FAQs) are highly recommended. This would not only help save time when finding lessons across the platform but would also offer motivation to sit through the lessons.

## 7.3 Implications for professional working mothers

Although online MBA programmes permit greater flexibility in terms of learning time and space, which potentially allows Chinese professional working mothers to fulfil multiple roles, the traditional time constraint arising from the uneven division of domestic responsibilities and caring remains. This finding supports previous recommendations that it is necessary to seek wider systems of social support for balancing work-family demands and career aspirations, including having supportive managers and support networks outside the workplace, flexible work hours, access to challenging assignments, and influential decision makers and having clearly defined requirements for advancement and career paths (Ibeh et al, 2008).

## 7.4 Limitations

This study has the following limitation. While the use of the snowballing sampling method proved most suitable for this study, it introduced bias, as the method reduces the likelihood that the sample will represent a good cross-section of the population. This shortcoming was manifested in this research by the predominance of students from science backgrounds. Thus, these findings may not necessarily be applicable to women across other professional fields, such as in the social sciences or arts.

# 8. Conclusions

The findings of this study confirm that work-family conflict is a significant problem for many female career aspirants. Many women are faced with the demands of performing multiple roles, such as employee/employer, wife, mother, a daughter to elderly parents, and student. Overall, the findings indicate a positive attitude toward online MBA programmes as a viable tool for facilitating work-family balance and career aspirations and a wealthy source of information. However, students' motivation for online learning, lack of networking provision, and perceived lack of quality assessment require consideration.

Although e-learning is already widely used in work settings and continuing education in China, it still lags behind in higher education, which is traditionally a conservative institution. Thus, China's societal culture has not yet fully accepted e-learning as an alternative to in-classroom education vis-à-vis its popularity in other parts of the world with established higher education e-learning, such as in the United Kingdom (the Open University), Spain (the Universidad Nacional de Educació́n a Distancia), or the Korean National Open University. It should be noted, however, that the transition from traditional classroom learning to e-learning cannot occur instantaneously, requiring time for the users to adjust (Arbaugh, 2004). This issue might be perceived differently by the cyber generation, who have never lived in a world without cyber technology, which needs further research.

This paper is intended to serve as a timely reminder to Chinese business schools and other key stakeholders of the need to revisit discussions of gender-supporting policies and innovative e-learning delivery methods tailored to the needs of professional working mothers in higher education, a group constantly overlooked by policymakers. In opening these discussions, higher education and companies will be addressing the dire issue of the continuing under-representation of women in business schools and managerial positions. This paper provides a fertile field for scholars to develop future theories and advance research and knowledge.

## Acknowledgments

The authors would like to thank the anonymous reviewers for their valuable suggestions and constructive comments. This work was supported by the Fundamental Research Funds for the Central Universities (2012QN208-HUST) and a grant from the Modern Information Management Research Center at Huazhong University of Science and Technology.